\documentclass{aa}
\usepackage{graphicx}
\usepackage{amsmath}

\begin{document}
\renewcommand{\theequation}{\thesection.\arabic{equation}}
\begin{titlepage}

\titlerunning{Microlensing towards M31 with MDM data}
\title{Microlensing towards M31 with MDM data}
\authorrunning{S. Calchi Novati et al.}
\author{S. Calchi Novati$^{1}$, Ph. Jetzer$^{1}$, G. Scarpetta$^{2}$,
Y. Giraud-H{\'e}raud$^{3}$, J. Kaplan$^{3}$, S. Paulin-Henriksson$^{3}$, A. Gould$^{4}$}

\offprints{novati@physik.unizh.ch} 
\institute {
$^{1}$ Institute of Theoretical Physics, University of Z{\"u}rich, Winterthurestrasse 190, 8057 Z{\"u}rich, Switzerland\\
$^{2}$ Dipartimento di Fisica ``E.R. Caianiello'', Universit{\`a} di Salerno, Via S. Allende, 84081 Baronissi (Sa),  
and INFN Sez. di Napoli - Gruppo Collegato di Salerno, Italy\\
$^{3}$ Physique Corpusculaire et Cosmologie, Coll{\`e}ge de France, 11 place M. Berthelot, 75231 Paris, France\\
$^{4}$ Department of Astronomy, Ohio State University, Columbus, OH 43210, USA
}

\date{Received/ Accepted}
\maketitle

\begin{abstract}
We report the final analysis of a search for 
microlensing events in the direction of the Andromeda galaxy,
which aimed to probe the MACHO composition of the M31 halo
using data collected during the 1998-99 observational campaign at the MDM observatory.
In a previous paper, we discussed the results from a first set of observations.
Here, we deal with the complete data set, and we take
advantage of some INT observations in the 1999-2000 seasons.
This merging of data sets taken by different instruments turns
out to be very useful, the study of the longer baseline available 
allowing us to test the uniqueness characteristic
of microlensing events. As a result, all the candidate
microlensing events previously reported turn out to be 
variable stars. We further discuss a selection based on different criteria,
aimed at the detection of short--duration events. We find three candidates
whose positions are consistent with self--lensing events,
although the available data do not allow us to conclude
unambiguously that they are due to microlensing.

\end{abstract}
 \vfill
\end{titlepage}

\section{Introduction} \label{intro}
Gravitational microlensing is a powerful tool
for the detection of dark matter in galactic haloes
in form of MACHOs (\cite{pacz86}). 
Intensive searches in our Galaxy have shown
that up to 20\% of the halo could be formed by objects
of around $M \sim 0.4 M_{\odot}$ (\cite{macho2000,eros2000}). 
This result is still debated (e.g. \cite{jetzer02}), and remains to be confirmed.
For this purpose it is useful
to  probe the MACHO distribution along different lines of sight.

A survey of M31, nearby and similar to our own Galaxy,
can supply several insights into this problem. Briefly,
it tests a different line of sight in our Galaxy, 
it allows to probe M31's own halo globally
and finally, the inclination of the M31 disk should give
an unmistakable signature of microlensing events (\cite{crotts92,agape93,jetzer94}).

Several collaborations have undertaken a search for microlensing
events towards M31 in the past years: AGAPE (\cite{agape99});
MEGA (\cite{crotts00}); POINT-AGAPE (\cite{point01}); 
WeCapp (\cite{wecapp}). 
In this same framework we have already reported on the analysis of a partial set of the MDM data 
in Calchi Novati et al. (2002, hereafter \cite{mdm02}). 
Furthermore, the POINT-AGAPE collaboration has given evidence of 4 
short high S/N ratio microlensing candidates (\cite{point02}, 2003).

In the present paper we conclude the analysis of the MDM data started in \cite{mdm02}.
We complement the MDM data with INT observations taken by the POINT-AGAPE collaboration.
Furthermore, we discuss the issue of selection criteria for
microlensing events, especially with respect to the background
of variable stars, and we present results from an analysis based 
on different criteria. In $\S$\ref{setup} we summarize the observational setup
and give some details of the data analysis, dealing in particular with some improvements
in the selection procedure; in $\S$\ref{int} we introduce the INT data
and discuss how we use the longer baseline they give us to construct a strong
supplementary selection criterion; $\S$\ref{selezione} is devoted to
the results of the analysis; in $\S$\ref{end} we present our conclusions.

\section{Observations and data analysis} \label{setup}

\subsection{Observational setup}

We analyse data collected on the 1.3 m McGraw-Hill
Telescope, at the MDM observatory, Kitt Peak (USA),
towards the two sides of M31 including the bulge\footnote{The
data are shared with the MEGA collaboration.}.
Two fields are observed, located in
$\alpha=$ 00h 43m 24s, $\delta  = 41^{\!\circ}12'10''$
(J2000) (``Target'', whose data have been the object
of the analysis presented in \cite{mdm02}) and $\alpha=$ 00h 42m 14s,
$\delta  =41^{\!\circ}24'20''$ (J2000) (``Control''). 
The Target field is centred on the far side of M31 while
the Control field is centred on the near side.
Throughout the observations a $2048 \times 2048$ pixel
CCD camera, with a field of view of $17'\times 17'$, has been used.

Two filters, similar to standard $R$ and $I$ Cousins,
have been used in order to test achromaticity. 
Furthermore, this colour information gives us
the chance to have a better check on red
variable stars, which can contaminate the search
for microlensing events.

We analyse data taken starting from October 1998 to the end of
December 1999. While the baseline of the Control field is the same as that of the Target field, 
the Control field has only about 20 nights of observations for both filters\footnote{Each 
night $\sim$ 20(11) images are taken in the $R$($I$) filter, which are 
then averaged to get the single image per night that we use in the following analysis.}, 
which is only about half as much data as  the Target field.
The average value of the seeing is $\sim 1.\hskip-2pt ''6$.

The photometric calibration is done with respect to a sample
of reference secondaries identified in our frame (\cite{magnier93}). 
As the photometric conditions for the reference images in the two fields are similar,
the zero point calibrations reported in \cite{mdm02} still hold.

\subsection{Reduction and analysis}

The reduction procedure and the candidate selection were discussed
in detail in \cite{mdm02}. Here we just recall the basic points
and discuss in more detail some aspects that have since been improved.

\subsubsection{Pixel Lensing}

Due to the distance of the target, the potential sources of microlensing events
are not resolved stars. We use the pixel lensing technique developed
by the AGAPE group to detect flux variations of unresolved sources 
(\cite{agape97})\footnote{An alternative method based on image subtraction is currently used
by the MEGA collaboration (\cite{tc96}).}. 
In order to cope with photometric and seeing variations, we first choose
for each filter a reference image (the $R$ reference image is also the geometric reference 
for both filters) and then calibrate the flux\footnote{With the notable
exception of geometrical alignment, throughout the analysis we substitute
for the pixel value the sum of the fluxes taken in a square of $5\times 5$
pixels around the central pixel, i.e., the corresponding \emph{superpixel}
value.} of all other images with respect to the reference image by means of a linear correction.
Our seeing correction is  empirical
and does not require the evaluation of the PSF of the image (\cite{mdm02}).

\subsubsection{Bump detection} \label{bumpnew}

Following \cite{mdm02}, we look for  bumps (in $R$ band) by a statistical analysis
of the light curve. To this end we construct the two estimators
for the significance of a variation $L$ and $Q$. We define 
\begin{equation} \label{likelihood}
L = -\ln\left(\Pi_{j\in bump}P(\Phi|\Phi>\Phi_{j})\right)
\;\;\mbox{given}\;\; \bar{\Phi}_{bkg},\, \sigma_{j}\,,
\end{equation}
where
\begin{equation}
P(\Phi|\Phi>\Phi_{j}) =  \int_{\Phi_{j}}^\infty d\Phi {\frac{1}{
\sigma_{j}\sqrt {2\pi}}} \exp\left[{-\frac{(\Phi
-\bar{\Phi}_{bkg})^2}{2\sigma_{j}^{2}}}\right];
\end{equation}
$\Phi_j$ and $\sigma_j$ are the flux and its associated error 
in a superpixel at time $t_j$, $\bar{\Phi}_{bkg}$ is an estimator of the
baseline level, and a ``bump'' is defined as a variation with 
at least 3 consecutive points more than $3\,\sigma$ above the baseline.  We define
\begin{equation}\label{rapp}
  Q\equiv \frac{\chi^2_{const}-\chi^2_{pacz}}{\chi^2_{pacz}/\rm dof}\,,
\end{equation}
where $\chi^2_{const}$ is the $\chi^2$ calculated with respect to the 
constant-flux 
hypothesis and $\chi^2_{pacz}$ is the $\chi^2$ calculated with respect to a
Paczy\'nski fit. We stress that $Q$ is evaluated along the entire light curve,
while $L$ is evaluated only along the bump. 

In \cite{mdm02} we carried out the analysis to detect clusters
of variable light curves using $L$. 
We eventually retained from the sample of selected light curves
(the ones with the highest value of $L$ in a given cluster)
only those characterized by $Q>Q_{thresh}$.

Here we follow a somewhat different approach
in that we select clusters related to flux variations
by demanding $Q>Q_{thresh}$.
Compared to the selection based on $L$, we are biased
in favor of monobump-shape variations. 
Moreover, we exclude all those light curves
showing spurious variations for which a high value of $L$ 
can happen to  be induced, e.g.,
by an underestimation of the baseline level.

To conclude with the bump detection, we evaluate
for all the pixels in each cluster $L$, and we then proceed as in \cite{mdm02}.
This is a better approach because $L$ is based
on an evaluation of the baseline 
more appropriate to this purpose, that is, 
to selecting the light curve showing the maximum
flux deviation from the baseline.

\subsubsection{Shape analysis}

For microlensing events the flux variation must be unique,
it must follow (in the point-like and uniform--motion approximations) 
the symmetric Paczy\'nski shape, and it must be achromatic.

The flux as a function of time for an event with amplification $A(t)$
and an unlensed source $\phi^*$ can be written as
\begin{equation} \label{lightcurve}
\phi(t)=\phi_{bkg}+\left[A\left(t\right)-1\right]\phi^*,
\end{equation}
where $\phi_{bkg}$ is the background flux including $\phi^*$.

The achromatic shape analysis is carried out as in \cite{mdm02}.
We perform a 7-parameter Paczy\'nski fit to the two bands simultaneously (background level $\phi_{bkg}$,
source flux without amplification $\phi^*$, and the parameters of the amplification:
time of maximum amplification $t_0$, Einstein time $t_{\textrm E}$, and the impact parameter $u_{min}$).
In the absence of direct knowledge of the source flux from e.g., a 
\emph{Hubble Space Telescope (HST) image} (e.g., \cite{point01}) or an 
indirect measurement of it from a high signal-to-noise ratio lightcurve
(e.g. \cite{point03}), there is a degeneracy among
the three parameters $\phi^*$, $t_{\rm E}$, and $u_{min}$ (\cite{gould96}). 
For these cases we retain two ``effective parameters'', the time width
$t_{1/2}=t_{1/2}\left(t_{\textrm E},\,u_{min}\right)$
and the excess at maximum with respect to the background 
$\Delta \phi_{max} = \Delta \phi_{max}\left(\phi^{*},\,u_{min}\right)$
(from which we evaluate the ``magnitude at maximum'' $R_{max}$
and the colour at maximum $(R-I)_C$). 
For an event to be achromatic we have to test  whether, along the bump,
the ratio of the deviation of the flux from the background 
in the two bands remains
constant in time ($\Delta \phi_R/\Delta \phi_I=\phi^*_R/\phi^*_I=$ 
constant)\footnote{Note that this same test can be applied to 
resolved-source microlensing, provided that ``background'' is replaced
by ``baseline'', i.e. the combined flux from the true source plus any
unresolved blended companions.  That is, blending breaks achromaticity
only for the ratio of the total fluxes in two bands, not differences in
these fluxes from baseline.}.

Furthermore, we carry out the Durbin-Watson test (\cite{dw51}), 
which is sensitive to time correlation among consecutive residuals.
In effect, it is able to check whether 
consecutive points lie, for instance, all above
or below the best Paczy\'nski fit shape, such behaviour
being characteristic of variable stars
(and for which the $\chi^2$ test can say little).
The test is based upon the evaluation of a
Durbin-Watson coefficient, $dw$, which must lie in an interval that depends on the number of points
along the light curve, e.g. for 20 points it must be $1.41(1.15)<dw<2.59(2.85)$
at the 10\% (2\%) level of significance.

\section{The baseline of INT data as a test for bump uniqueness}
\label{int}

\subsection{The problem of variable sources} \label{variabili}

A main problem in the search for microlensing events is the estimate
of the background noise given by variable sources. This is particularly serious
in the case of pixel lensing for two main reasons. First,
the class of stars to which we are in principle most sensitive are the red giants, 
for which a large fraction are variable stars 
(regular or irregular). Second, as we look for \emph{pixel}
flux variations, it is always possible to collect (in the same pixel)
light from more than one source whose flux is varying.

Thus, in the analysis we are faced with two problems: 
large-amplitude variable sources
whose signal can mimic a microlensing signal,
and variable sources of smaller amplitude whose signal can 
give rise to non-Gaussian fluctuations
superimposed on the background or on other physical variations.

In \cite{mdm02} we followed a sufficiently conservative  approach to minimize
the impact of these problems. We adopted severe criteria
in the shape analysis with respect to the Paczy\'nski fit
(stringent cut for both the $\chi^2$ and the Durbin-Watson test)
and, furthermore, we eliminated candidates  with both a long timescale
($t_{1/2}>40$ days) and a red colour ($(R-I)_C>1$), since these most likely
originate from variable stars.

As we had already noted, this analysis suffered from
the intrinsic limitation of an insufficient baseline,
so that the variable-source issue could not be conclusively tested.

Here we take advantage of the opportunity afforded by access
to reduced INT data to extend our selected light curves. 
As we show, this will be a key element of our analysis.

\subsection{INT extension of MDM light curves} \label{prolungamento}

The INT fields almost completely cover the MDM fields (except for a sizable fraction 
of the Control field far from the M31 bulge and a narrow band in the Target field). 
The good sampling during the two INT seasons 1999-2000
of data acquired in Sloan $r'$ and $i'$ bands\footnote{For this analysis
we do not use the available $g'$ band data.}, together with their high quality,
allows a straightforward and quantitative analysis. 

As a first step we have calculated the astrometric\footnote{The INT pixel size
is $0.\hskip-2pt ''33$ versus $0.\hskip-2pt ''50$ for MDM pixels.} transformations
between the different fields (data from 4 different INT CCDs are needed to cover
the two MDM fields). We found that a global transformation
over one entire field led to systematic errors as large as $5''$. Hence, we 
use our large samples of 1194 and 845 references stars in the Target and
Control field, respectively, to make a linear (i.e., 6-parameter) local transformation 
for each selected pixel. We demand a minimum of 24 reference stars, which 
are usually found within a square of $\sim 2'-3'$ about the selected pixel. 
We thereby reach an average astrometric precision of $\sim 0.\hskip-2pt ''2$.
To this we add the error in locating the center
of the variation, so that on average we get $\sigma \sim 0.\hskip-2pt ''5$.

As a second step we have determined from the cross-analysis of
about 30 selected resolved stars in each field, 
a ``colour equation'' that allows us to align the MDM ($R$ and $I$)
and INT ($r'$ and $i'$) instrumental magnitudes closely
enough to permit a quantitative comparison of
the relative flux variations in each light curve.

The uniqueness condition for a given variation requires that the extension
of the light curve into the INT data should remain flat.
However, the analysis of the PA-N1 event (\cite{point01})  
has shown that a microlensing light curve may be contaminated 
by a neighbouring variable star. Therefore,
care must be taken before rejecting a microlensing event
because of a second bump, and it is essential to test the following issues:
the relative astrometric position of the two bumps
and the similarity of the shape of the two deviations.

Given a candidate microlensing light curve in the MDM data, 
we  locate the corresponding INT light curve,
and then calculate the estimator $L$ in a square of $7\times 7$ pixels
around the central pixel to check whether
there is a variation, and where it is located. If 
we detect a variation in the INT data,
we calculate its amplitude and width with a Paczy\'nski fit in order to
compare quantitatively the two variations.

As a criterion of rejection for a given
MDM candidate microlensing event for which
we find a variation on its extension into
the INT data, \emph{i.e.}, a likely variable star,
we demand that the position of the INT variation be compatible 
with the corresponding MDM variation
within $3\,\sigma$ \emph{and} that the relative widths and deviations of flux
lie within $6\,\sigma$ as evaluated from the two independent
Paczy\'nski fits\footnote{We allow here a larger margin
because we are aware that probing such an effect
can be difficult, especially when
dealing with variations showing both a long 
time width and a red colour,
these being possibly due to red variable
stars that do not necessarily show a strictly
periodic and regular behaviour.}.

\section{Candidate selection} \label{selezione}

\subsection{Target and Control fields: the first selection}  \label{first}

We  now present the results in both MDM  fields, Target and Control,
of an analysis complemented where possible
by the stability test using the INT data.

The threshold value for the estimators
of the significance of the bump are fixed as in \cite{mdm02}
($Q>100$ and $L_1>100$, but now we proceed
as explained in $\S$\ref{bumpnew}\footnote{We note that, in principle,
for a cluster of pixels with $Q>100$, we can have $L_1<100$.}). 
The selection criteria are the same as in \cite{mdm02}:
the selected light curves must have enough points along the bump (at least 4 points
on both sides of the maximum, and 3 inside the interval $t_0\pm t_{1/2}/2$);
for the Paczy\'nski fit we require that $\chi^2/\textrm{dof}<1.5$ and that the Durbin-Watson
test  be satisfied at the 10\% level; we require that \emph{either}
$t_{1/2}<40$ days \emph{or} $(R-I)_C<1$. 

Due to the different approach we follow in bump detection, 
in the Target field we find 1 more light curve in addition to the 5 we already reported in \cite{mdm02}.
In the Control field we find 4 light curves 
compatible with microlensing.

\begin{figure*}
\resizebox{7 cm}{!}
{\includegraphics{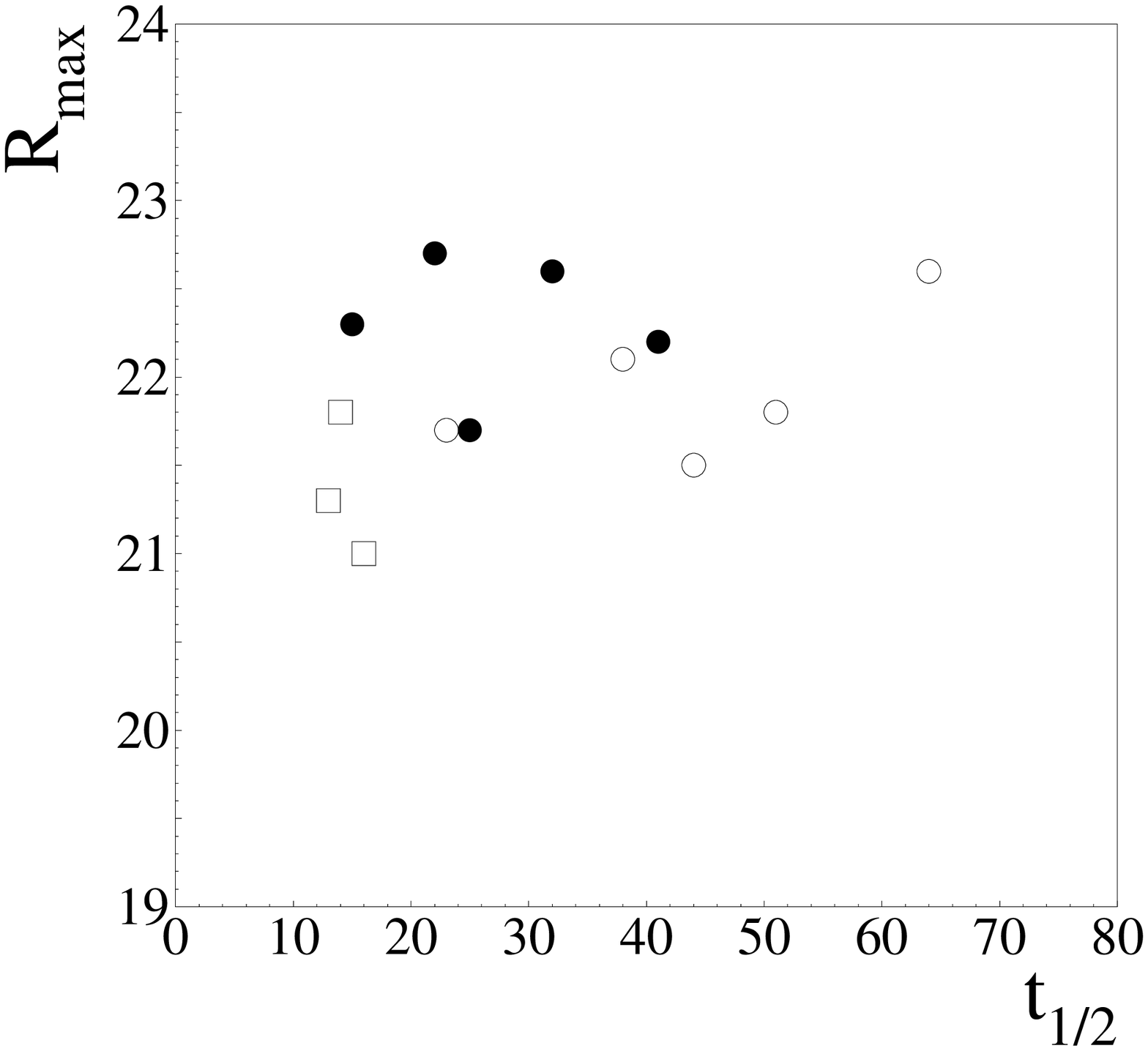}}
\resizebox{7 cm}{!}
{\includegraphics{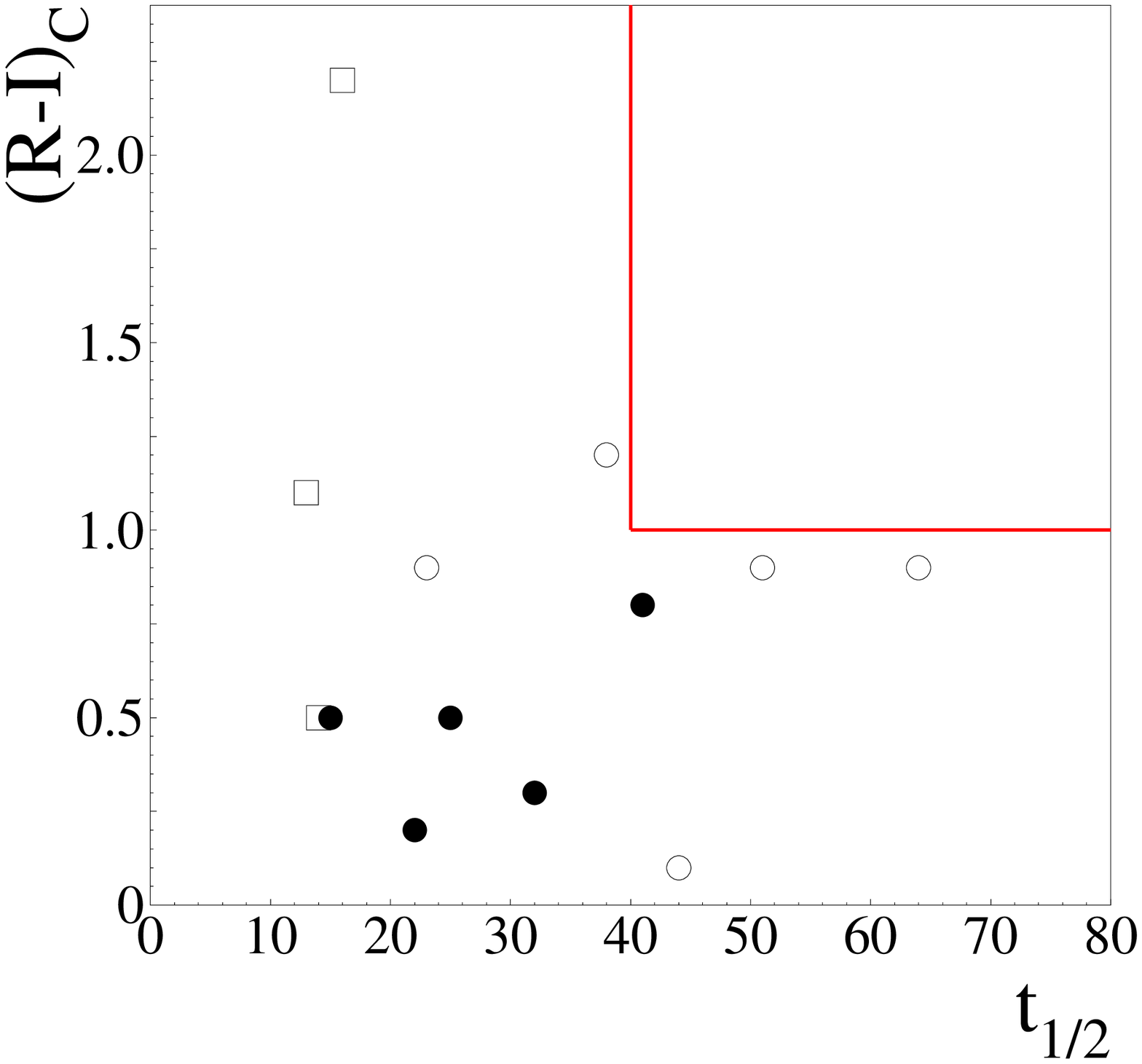}}
{\caption{Physical characteristics for the selected light curves in $\S$\ref{first} (circles). 
Filled symbols represent candidates previously reported in Target field. Squares represent 
the result of the analysis presented in $\S$\ref{new}. In the right panel, 
the lines at $t_{1/2}=40$ days and $(R-I)_C=1$ indicate our excluded region of this parameter space.}
\label{plot1}}
\end{figure*}

We summarize the main physical characteristics of these
10 events in Fig. \ref{plot1}, which shows
$R_{max}$ vs $t_{1/2}$ and $(R-I)_C$ vs $t_{1/2}$.


As can be noted, with respect to the selected light curves
in \cite{mdm02}, we tend to lack short timescale candidates among the new events and, 
moreover, 3 of the 5 candidates lie at the boundary
we have fixed for the colour-timescale compatibility  with a microlensing signal.

The INT data allow us to check the stability of all 5 Target
light curves and 2 of the 4 Control light curves.
In Table \ref{tavola1} we summarize the results of this
comparative analysis, relative astrometry and  relative shape analysis.

\begin{figure*}
\resizebox{7 cm}{!}
{\includegraphics{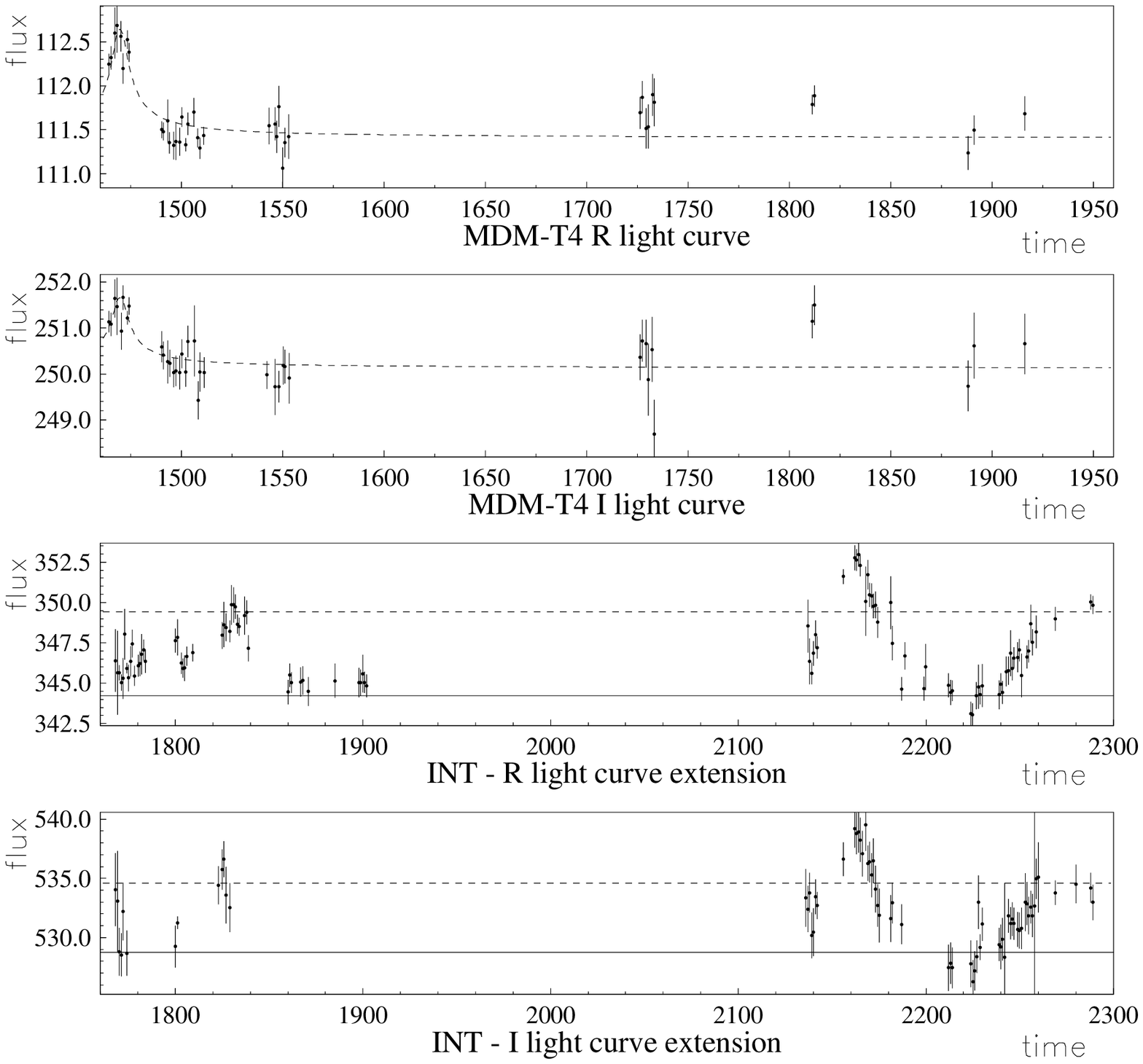}}
\resizebox{7 cm}{!}
{\includegraphics{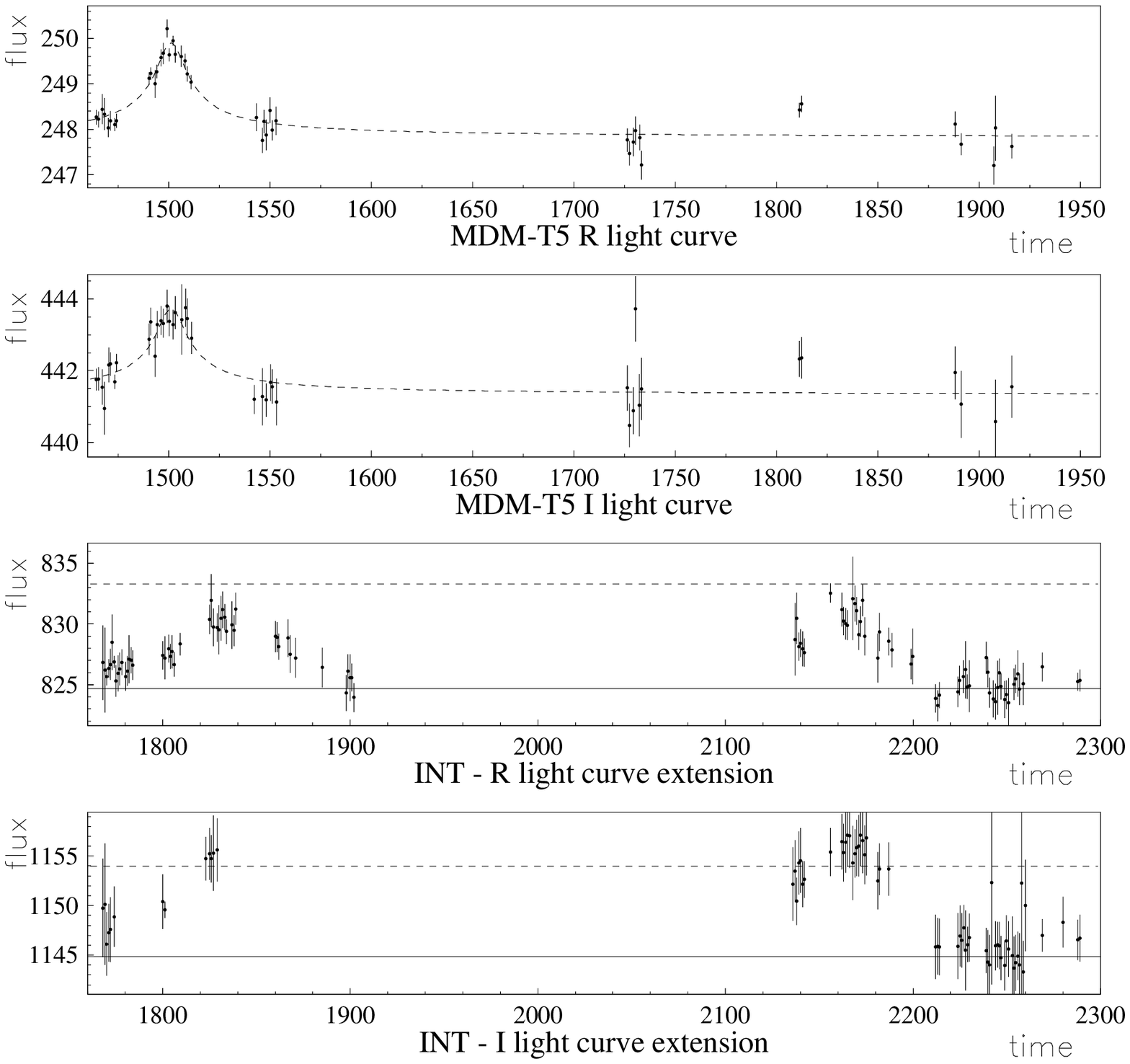}}
{\caption{MDM T4 (left) and T5 light curves together with their extensions into the INT data.
On the $y$ axis, flux is in ADU/s; on the $x$ axis, time is in days, with the origin in J-2449624.5 (both data sets). 
For the MDM light curves the dashed line represent the result of the Paczy\'nski fit.
For the INT light curves, shown together with the solid line representing
the baseline is a dashed line representing the level of the maximum deviation of flux found on the corresponding MDM light curve.}
\label{plot2}}
\end{figure*}

As can be seen all  8 MDM events show compatible variations
on their INT extensions. (In particular, we note that the positions of the two bumps
are always compatible at $1\,\sigma$ level).
As a result, \emph{all} the checked light curves
are \emph{rejected} as possible microlensing candidates.
As an example, we show MDM light curves T4 and T5 (respectively,
the shortest and the brightest flux variations detected) together with their INT extensions (Fig. \ref{plot2}).


This analysis shows how the sample of microlensing candidates
derived from this selection is strongly contaminated by variable stars.
The two additional Control light curves for which there are no INT extensions 
have parameters: $t_{1/2}=51\pm 5,\,64\pm 6$ days and $(R-I)_C=0.9\pm 0.2$,
i.e., they are relatively long candidates. It then follows that,
without INT vetting, it is no longer reasonable to look at them
as viable microlensing candidates. Since, also looking at the results
of the previous analysis, we suspect them
to be red variable stars but cannot prove this without an extended
baseline, we must exclude, when searching for such long variations,
the non-INT portions of the MDM fields from the analysis.

\subsection{Search for short duration events} \label{new}

The analysis carried out in the previous section shows 
that variable sources can mimic the Paczy\'nski shape
quite well, even with data available in two bands.
It seems therefore appropriate in the search
for viable microlensing candidates to relax 
the criteria introduced to characterize
the shape of the variation and, on the other hand,
to restrict the allowed space of physical parameters.
We remark that the longer baseline now available make
this approach viable.
This search is also motivated by the lack 
of self--lensing events, which we would expect to find.

We then proceed to a new selection with the same threshold values
for $L$ and $Q$. For the temporal sampling, we require at least 3 points
inside the interval $t_0 \pm t_{1/2}/2$, with at least one
on each side of $t_0$. We then impose the following conditions:
\begin{itemize}
\item $\chi^2/\textrm{dof}<5$;
\item DW test at $2\%$ significance level;
\item $t_{1/2}<20$ days\footnote{We note that these short timescales are
consistent with what we expect from the Monte Carlo
simulations discussed in \cite{mdm02}.}.
\end{itemize}

This search yields 8 additional light curves (4 in each
field) which have
durations $t_{1/2} \in (13,20)$ days
and flux deviations $R_{max} \in (21.0,22.8)$. 
The stability test on the INT
data, as outlined in $\S$\ref{prolungamento},  
allows us to reject 5 of these variations
as microlensing candidates.
We are left with three light curves, all lying in the Control field.
We label these C3, C4 and C5, the last having no INT extension.
The issue of the stability on the INT extension of the two remaining 
candidates deserves some additional comments. 

MDM-C3: An INT variation is detected only in the $i'$ light curve.
Its position is compatible within $1\,\sigma$ to that
of the MDM variation. We note that the observed
$i'$ variation is significantly smaller
than that observed on the MDM $I$ light curve
and, as mentioned, the $r'$ light curve is flat.
Moreover, the time width
of the MDM variation is actually quite short.
This analysis indicates that, although sitting
in the same position, the two variations may be
due to different sources. At the same time
we are aware that the INT data do not  allow us 
to fully characterize the shape of the bump.
We thus consider the stability test
for this candidate to be \emph{inconclusive}.

MDM-C4: We detect a variation on the INT data
that we localize within $1\,\sigma$ 
from the corresponding MDM variation.
We evaluate the variations of flux in $r'$ and $i'$
as being compatible within 3 and $1\,\sigma$ 
respectively with the corresponding variations
on the MDM light curve. However, the observed
time width of the INT variation, $t_{1/2}\sim 70$ days,
is significantly larger than the evaluated
time width of the MDM variation, $t_{1/2}\sim 13$ days.
As is the case for C3, the shape analysis may indicate
that we are observing on the same light curve
a variable star (a likely long period red variable)
\emph{and} a microlensing event. 
Our data are, however, insufficient to confirm 
or to reject this hypothesis.
In this case too we then consider the stability test
for this candidate to be \emph{inconclusive}.

In Table \ref{tavola2} we report the main characteristics of these
light curves, position and physical parameters
as evaluated from the Paczy\'nski fit. We note that, 
lacking any information from other sources (e.g. \emph{HST} images)
to measure the flux of the unamplified source,
and since, on the basis of the fit alone, the
data do not allow us to break the parameter degeneracy,  
we have no way to get any reliable 
information on the physical parameter $t_{\textrm E}$,
so that also no reliable estimate of the mass of the lens is possible.



\begin{figure*}
\resizebox{7 cm}{!}
{\includegraphics{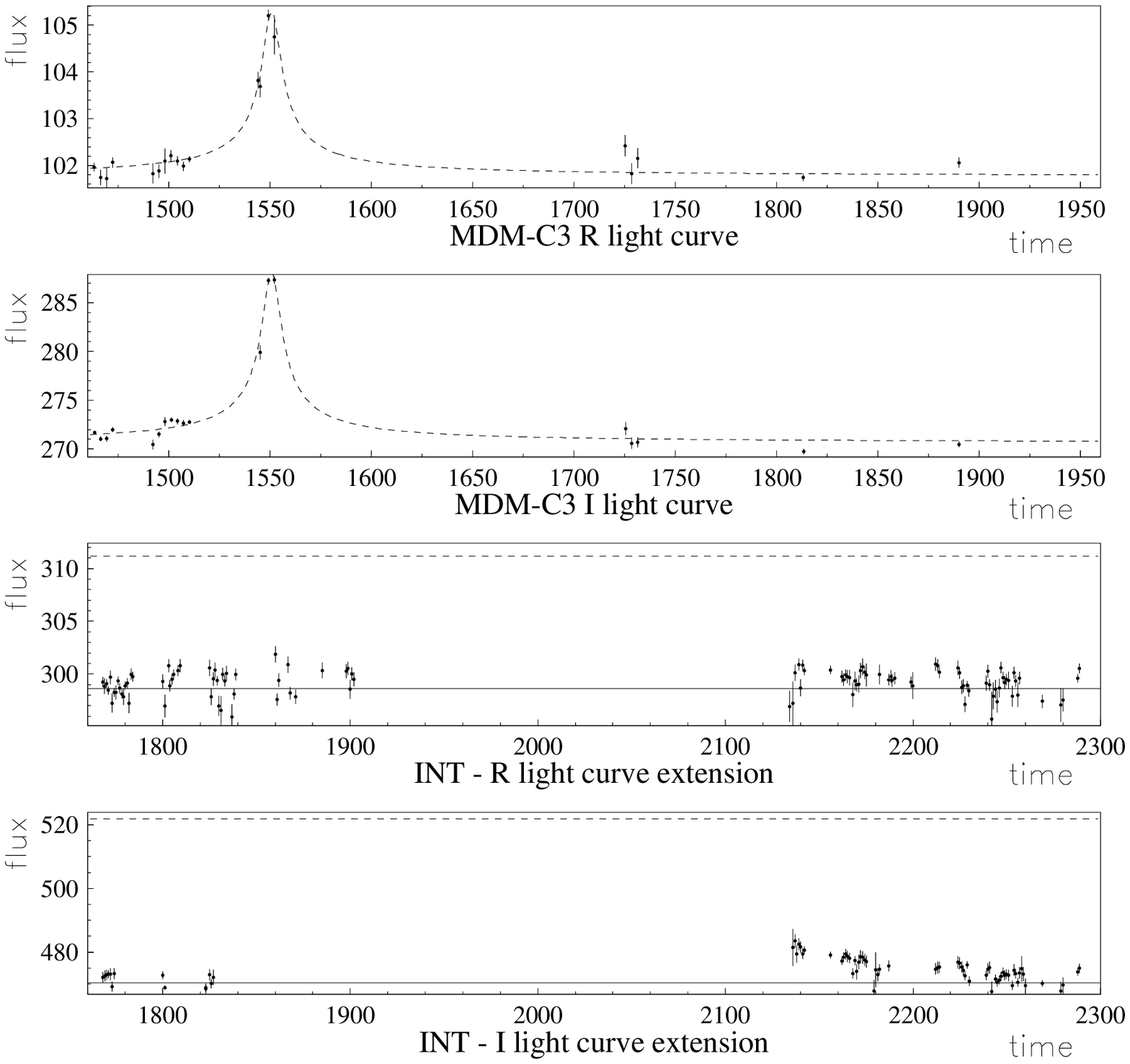}}
{\caption{The MDM-C3 light curve (top) and its corresponding INT extension. Notation
is the same as in Fig. \ref{plot2}.}
\label{cl1}}
\end{figure*}
\begin{figure*}
\resizebox{7 cm}{!}
{\includegraphics{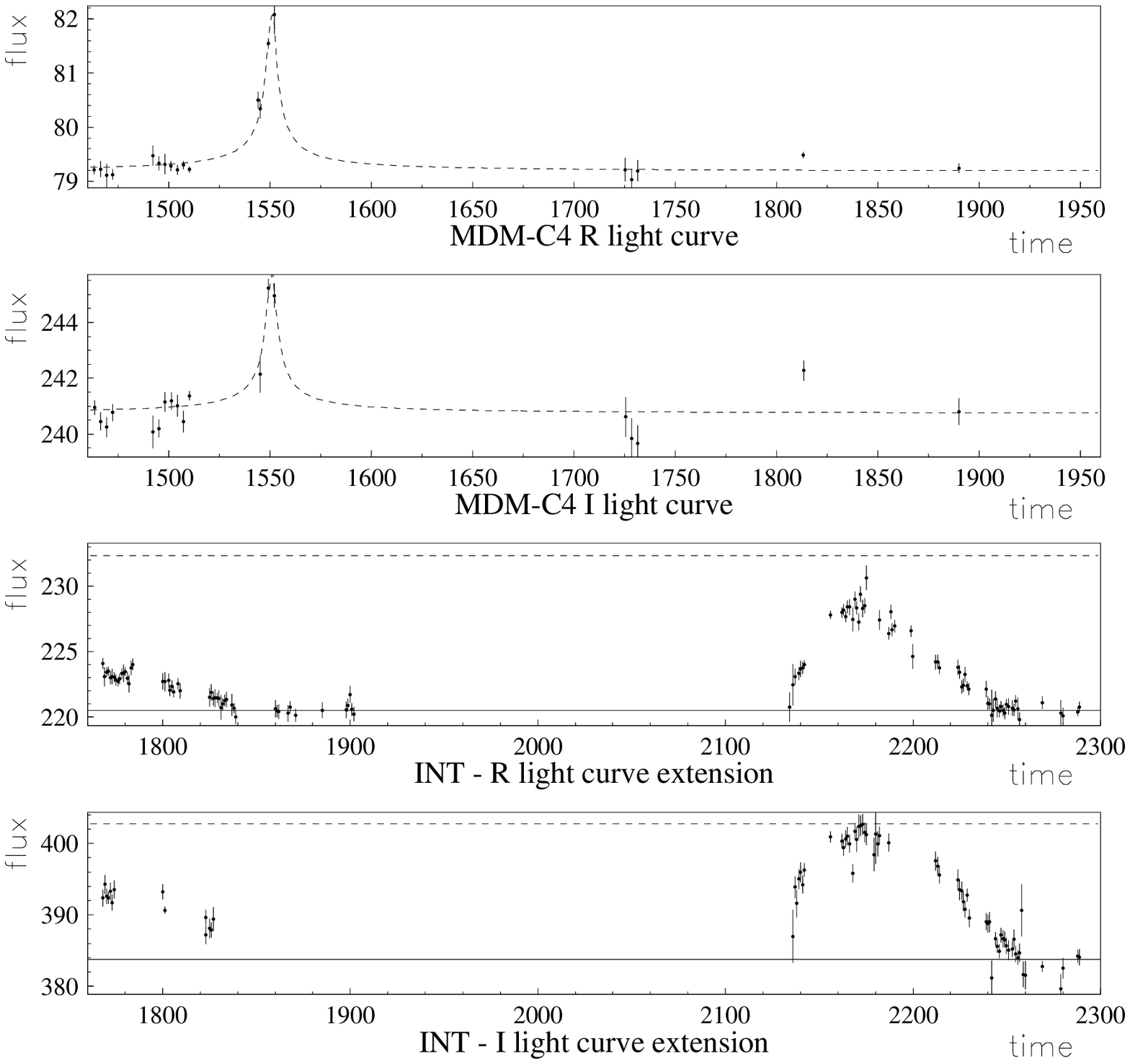}}
{\caption{The MDM-C4 light curve (top) and its corresponding INT extension. Notation
is the same as in Fig. \ref{plot2}.}
\label{cl2}}
\end{figure*}
\begin{figure*}
\resizebox{7 cm}{!}
{\includegraphics{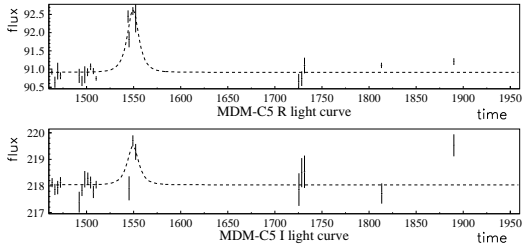}}
{\caption{The MDM-C5 light curve, for which no extension on INT data is available. 
Notation is the same as in Fig. \ref{plot2}.}
\label{cl3}}
\end{figure*}

As can be noted by looking at the light curves (Figs. 
\ref{cl1}, \ref{cl2} and  \ref{cl3}),  
the sampling does not allow
us to test  conclusively the symmetry of the bump.
Regarding the achromaticity, the ratio of $R$ and $I$  deviations
from the baseline for the points belonging
to the bump is about constant, as expected for microlensing events.
We note that the MDM-C3 light curve, 
whose variation is quite short and very red, occupies
a peculiar position in the $(R-I)_C$--$t_{1/2}$ parameter space
(Fig. \ref{plot1}, upper left of right panel).

Even if the available data do not allow us to draw 
firm conclusions on the nature of the selected flux variations,
their projected distance from the centre of M31 (see Table \ref{tavola1})
make them marginally consistent with the self--lensing events that
we expect according to the Monte Carlo simulations discussed in \cite{mdm02}.


\subsection{Comparison with POINT-AGAPE results}

In the framework of microlensing searches towards M31,
the most promising results have been so far reported
by the POINT-AGAPE collaboration. In particular,
they have identified 4 high S/N (likely to be microlensing) 
light curves with high flux variation,
$R_{max}<21$, and short timescale ($t_{1/2}<25$ days,  with
3 that are as small as $t_{1/2} \sim 2$ days), 
two of which lie quite near the M31 bulge,
where, however, data still do not allow one to answer unambiguously the question of whether
they are due to self--lensing effects or to MACHOs (\cite{point03}).
We note that we lack, in the MDM data, such short duration
and highly amplified events. We conjecture that this apparent discrepancy 
can plausibly be explained by the relative quality of the data sets 
In particular, relative to the MDM data set,
the INT data cover a much ($>3$ times) larger portion of the sky
around the M31 bulge, with more than the double the nights of observation.
In particular, in the MDM Control field, with at best one point every
3 nights, we can obviously not characterize events of such short duration. 
On the other hand, we point out that while the POINT-AGAPE analysis
has been carried out with a threshold on the magnitude at maximum
of the detected variations ($R_{max}<21$), in the analysis
presented here we do not adopt such a threshold
(we recall, see \cite{mdm02}, that we are sensitive to variations
down to around $R_{max}\sim 23$). 

\section{Discussion and perspectives} \label{end}

We have reported on the status of the analysis of the MDM data
carried out in the framework of a search for microlensing events
towards M31. As a result of a full analysis of 2 years of data, in which
we have also taken advantage of the longer baseline made available
by the use of some INT data, we have first excluded as viable
all the microlensing candidates previously reported in \cite{mdm02}.
Second, we have reported the selection of three more candidates
compatible with a Paczy\'nski light curve, which may possibly be 
due to self lensing.  
On examining their extensions into the INT data, we find a second
variation in both cases for which such data are available.  However,
these variations are strikingly different from and much longer than
the short-duration events seen in the MDM data.  Hence, it is  
plausible that each of these second variations is due to an unrelated
variable star superimposed on a true microlensing signal.  The
available data do not permit us to resolve this question.

We have discussed the issue of variable-star contamination of the signal:
we note that this is a significantly more serious problem than in the cases of 
the search for microlensing events towards the Magellanic clouds and the 
Galactic bulge in which one monitors the flux variations of resolved sources 
of known type.

The analysis reported in this paper shows once again 
the crucial role played by a frequent sampling of the data and the total baseline length,
both of which are essential to getting meaningful results when extracting microlensing signals
from the background of variable sources. 
In this light, the forthcoming full analysis of the detection efficiency, 
together with results from the new sets of data acquired in the 2001 and 2002 seasons
at both 1.3m and 2.4m MDM telescopes with a new 8K-CCD array  
(as well as the prospective 2003 campaign at the TT1 telescope
in the south of Italy, \cite{slott}), should eventually give us the opportunity
to draw firmer conclusions on the issue
of the MACHO fraction in the halo of M31 galaxy.

\begin{acknowledgements}
We thank the referee for useful comments and suggestions.
We are grateful to the POINT-AGAPE collaboration for allowing
us the access to their data set.
SCN was supported by the Swiss National Science Foundation
and by the Tomalla Foundation. AG was supported by grant 02-01266 from the US NSF.
\end{acknowledgements}

\newpage

\begin{table*}
\begin{tabular}{|c|c|c|c|c|}
\hline
id & $\sigma$(position) & $\sigma(t_{1/2})$ & $\sigma(\Delta\Phi_R)$ & $\sigma(\Delta\Phi_I)$  \\
\hline
T1 & 1 & 1 & 2 & 2 \\
T2 & 1 & 1 & 1 & 2 \\
T3 & 1 & 4 & 4 & 3 \\
T4 & 1 & 4 & 5 & 3 \\
T5 & 1 & 3 & 2 & 3 \\
T6 & 1 & 2 & 1 & 1  \\    
\hline
C1 & 1 & 1 & 2 & 4 \\
C2 & 1 & 2 & 2 & 2\\
\hline
\end{tabular}
\caption{Results of the stability analysis on the INT data extension
of the selected MDM microlensing candidates. For each MDM light curve (T and C stand
respectively for Target, where we use the same numeration used in \cite{mdm02},
and Control) for which we find an astrometrically compatible INT bump, we report  
the number of standard deviation within which the position and 
the bump parameters (duration and flux deviation in both bands
as calculated from independent Paczy\'nski fits on each of the MDM and INT light curves), 
of the two variations are compatible.}
\label{tavola1}
\end{table*}

\begin{table*}
\begin{tabular}{|c|c|c|c|c|}
\hline
&& C3 & C4 & C5\\
\hline
\hline
$\alpha$ (J2000) && $00^h\,42^m\,12.5^s$ & $00^h\,42^m\,25.6^s$ & $00^h\,41^m\,52.8^s$ \\
$\delta$ (J2000) && $41^\circ\, 21'\, 30''$ & $41^\circ\, 26'\, 27''$ & $41^\circ\, 17'\, 18''$ \\
$d$ && $8'01''$ & $10'53''$ & $9'45''$ \\
\hline
$\chi^2/\textrm{dof}$ && 2.90 & 2.32 & 2.47 \\
$dw_R$ && 2.48 & 2.01 & 1.78 \\
$dw_I$ && 1.62 & 2.23 & 1.92 \\
\hline
$t_{1/2}$ (days) && $16 \pm 2$ & $13 \pm 2$ & $14\pm 2$ \\
$R_{max}$ && $21.0 \pm 0.1$ & $21.3 \pm 0.1$ & $21.8\pm 0.1$ \\
$R-I$ && $2.2\pm 0.2$ & $1.1 \pm 0.2$ & $0.5\pm 0.2$\\
\hline
INT data &&  inconclusive & inconclusive & no extension\\
\hline
\end{tabular}
\caption{Main characteristics of the three selected short-event light curves; $d$
is the projected distance from the centre of M31.} \label{tavola2}
\end{table*}

\begin{thebibliography}{99}
\bibitem[Alcock et al. 2000]{macho2000} Alcock C.  et al. 2000, ApJ, 542, 281
\bibitem[Ansari et al. 1997]{agape97}  Ansari R. et al. 1997,  A\&A, 324,  843
\bibitem[Ansari et al. 1999]{agape99}  Ansari R. et al. 1999, A\&A, 344, L49
\bibitem[Auri\`ere et al.\ 2001]{point01}  Auri{\`e}re M. et al. 2001, ApJ, 553, L137
\bibitem[Baillon et al. 1993]{agape93} Baillon P., et al. 1993, A\&A, 277, 1
\bibitem[Bozza et al. 1999]{slott} Bozza V. et al. 1999, astro-ph/9907162
\bibitem[Paper I]{mdm02} Calchi Novati S. et al. 2002, A\&A, 381, 848 (Paper I)
\bibitem[Crotts 1992]{crotts92} Crotts A.P. 1992, ApJ, 399, L43
\bibitem[Crotts et al. 2000]{crotts00} Crotts A.P. et al. 2000, astro-ph/0006282
\bibitem[Durbin \& Watson 1951]{dw51} Durbin J. \& Watson G.S. 1951, Biometrika, 38, 159
\bibitem[Gould 1996]{gould96} Gould A. 1996, ApJ, 470, 201
\bibitem[Jetzer 1994]{jetzer94} Jetzer Ph. 1994, A\&A, 286, 426
\bibitem[Jetzer et al. 2002]{jetzer02} Jetzer Ph. et al. 2002, A\&A, 393, 129
\bibitem[Magnier et al. 1993]{magnier93}  Magnier E. A.  et al. 1993, A\&A, 272, 695
\bibitem[Lasserre et al. 2000]{eros2000} Lasserre T. et al. 2000, A\&A, 355, L39
\bibitem[Paczy\'nski 1986]{pacz86}  Paczy\'{n}ski B. 1986, ApJ, 304, 1
\bibitem[Paulin-Henriksson et al. 2002]{point02} Paulin-Henriksson et al. 2002, ApJ, 576, L121
\bibitem[Paulin-Henriksson et al. 2003]{point03} Paulin-Henriksson et al. 2003, A\&A, in press, astro-ph/0207025
\bibitem[Riffeser et al. 2001]{wecapp} Riffeser A. et al. 2001, A\&A, 379, 362
\bibitem[Tomaney \& Crotts 1996]{tc96} Tomaney A. \&  Crotts A. 1996, AJ, 112, 2872

\end{thebibliography}
\end{document}